\documentclass[prd,aps,twocolumn,nofootinbib,floats,letterpaper,floatfix,groupedaddress,eqsecnum]{revtex4}

\usepackage{amsmath}
\usepackage{amssymb}
\usepackage{color}
\usepackage{graphicx}
\usepackage{epstopdf}           
\usepackage{latexsym}
\usepackage{times}
\usepackage{ifpdf}

\usepackage{dcolumn,epsfig}
\usepackage{amssymb,amsmath}

\def\nn{\nonumber}

\def\be{\begin{equation}}
\def\ee{\end{equation}}
\def\beq{\begin{eqnarray}}
\def\eeq{\end{eqnarray}}

\def\IL{\relax{\rm I\kern-.18em L}}

\def\nn{\nonumber}

\begin{document}

\title{Breit-Wigner resonances and the quasinormal modes of
  anti-de Sitter black holes}

\author{Emanuele Berti} \email{berti@phy.olemiss.edu} \affiliation{Department
  of Physics and Astronomy, The University of Mississippi, University, MS
  38677-1848, USA}

\author{Vitor Cardoso} \email{vitor.cardoso@ist.utl.pt} \affiliation{Centro
  Multidisciplinar de Astrof\'{\i}sica - CENTRA, Dept. de F\'{\i}sica,
  Instituto Superior T\'ecnico, Av. Rovisco Pais 1, 1049-001 Lisboa, Portugal
  \\ Department of Physics and Astronomy, The University of Mississippi,
  University, MS 38677-1848, USA}

\author{Paolo Pani} \email{paolo.pani@ca.infn.it} \affiliation{Dipartimento di
  Fisica, Universit\`a di Cagliari, and INFN sezione di Cagliari, Cittadella
  Universitaria 09042 Monserrato, Italy \\Currently at Centro Multidisciplinar
  de Astrof\'{\i}sica - CENTRA, Dept. de F\'{\i}sica, Instituto Superior
  T\'ecnico, Av. Rovisco Pais 1, 1049-001 Lisboa, Portugal}

\begin{abstract}
The purpose of this short communication is to show that the theory of
Breit-Wigner resonances can be used as an efficient numerical tool to compute
black hole quasinormal modes. For illustration we focus on the Schwarzschild
anti-de Sitter (SAdS) spacetime. The resonance method is better suited to
small SAdS black holes than the traditional series expansion method, allowing
us to confirm that the damping timescale of small SAdS black holes for scalar
and gravitational fields is proportional to $r_{+}^{-2l-2}$, where $r_{+}$ is
the horizon radius. The proportionality coefficients are in good agreement
with analytic calculations.  We also examine the eikonal limit of SAdS
quasinormal modes, confirming quantitatively Festuccia and Liu's
\cite{Festuccia:2008zx} prediction of the existence of very long-lived modes
in asymptotically AdS spacetimes.  Our results are particularly relevant for
the AdS/CFT correspondence, since long-lived modes presumably dominate the
decay timescale of the perturbations.
\end{abstract}

\maketitle

\section{Introduction}

It is well known that quasi-bound states manifest themselves as poles in the
scattering matrix, and as Breit-Wigner resonances in the scattering
amplitude. Chandrasekhar and Ferrari made use of the form of the scattering
cross section near these resonances in their study of gravitational-wave
scattering by ultra-compact stars
\cite{ChandrasekharFerrari,Chandrasekhar:1992ey}. In geometrical units
($c=G=1$), the Regge-Wheeler potential $V(r)$ describing odd-parity
perturbations of a Schwarzschild black hole (BH) of mass $M$ has a peak at
$r\sim 3M$. Constant-density stellar models may have a radius $R/M<3$ (but
still larger than the Buchdal limit, $R/M>2.25$). When $R/M\lesssim 2.6$, the
radial potential describing odd-parity perturbations of the star (which
reduces to the Regge-Wheeler potential for $r>R$) displays a local minimum as
well as a maximum. If this minimum is sufficiently deep, quasi-stationary,
``trapped'' states can exist: gravitational waves can only leak out to
infinity by ``tunneling'' through the potential barrier. Since the damping
time of these modes is very long, Chandrasekhar and Ferrari dubbed them
``slowly damped'' modes \cite{ChandrasekharFerrari}.

For trapped modes of ultra-compact stars the asymptotic wave amplitude at
spatial infinity $\Psi \sim \alpha \cos \omega r+\beta \sin \omega r$ has a
Breit-Wigner-type behavior close to the resonance:
\be
\alpha^2+\beta^2\approx{\rm const}\,\left[(\omega-\omega_R)^2+\omega_I^2\right]\,,\label{breitwigner}
\ee
where $\omega_I^{-1}$ is the lifetime of the quasi-bound state and
$\omega_R^2$ its characteristic ``energy''.  
The example of ultra-compact stars shows that the search for weakly damped
quasinormal modes (QNMs) corresponding to quasi-bound states
($\omega=\omega_R-i\omega_I$ with $\omega_I\ll \omega_R$) is extremely
simplified. We locate the resonances by looking for minima of
$\alpha^2+\beta^2$ on the $\omega=\omega_R$ line, and the corresponding
damping time $\omega_I$ can then be obtained by a fit to a parabola around the
minimum \cite{ChandrasekharFerrari,Chandrasekhar:1992ey}.

Here we show that this ``resonance method'' can be used very successfully in
BH spacetimes.  The resonance method is particularly valuable in studies of
asymptotically AdS BHs.  The QNM spectrum of AdS BHs is related to
thermalization timescales in a dual conformal field theory (CFT), according to
the AdS/CFT conjecture \cite{Horowitz:1999jd}. Analytic studies of wave
scattering in AdS BHs have previously hinted at the existence of resonances
(see Fig.~9 in Ref.~\cite{Harmark:2007jy}); here we show that these are indeed
Breit-Wigner resonances.

Various analytic calculations recently predicted the existence of long-lived
modes in asymptotically AdS BH spacetimes
\cite{Grain:2006dg,Festuccia:2008zx,Daghigh:2008jz}.
These modes will presumably dominate the BH's response to perturbations, hence
the thermalization timescale in the dual CFT. Since their existence may be
very relevant for the AdS/CFT conjecture, we decided to investigate
numerically these long-lived modes. In Section \ref{small} we confirm the
existence of quasi-bound states for small SAdS BHs, first predicted by Grain
and Barrau \cite{Grain:2006dg}, partially correcting some of their
predictions.  In Section \ref{eikonal} we re-analyze the eikonal limit of SAdS
QNMs studied by Festuccia and Liu \cite{Festuccia:2008zx}, finding excellent
agreement with their calculations.

It may be useful to point out that a different but intimately related method
(the complex angular momentum approach, a close kin of the theory of Regge
poles in quantum mechanics) has been used in the past to study QNMs in
asymptotically flat BH spacetimes
\cite{Andersson:1994rk,Andersson:1994rm,Decanini:2002ha}. Some aspects of the
relation between the resonance method and the theory of Regge poles are
illustrated, for example, in Ref.~\cite{Chandrasekhar:1992ey}.

\section{\label{small}Quasi-bound states in SAdS black holes}

In this paper we will focus on SAdS BHs in four spacetime dimensions, but our
results are trivially extended to higher dimensions. Scalar ($s=0$),
electromagnetic ($s=1$) and vector-type (or Regge-Wheeler) gravitational
perturbations ($s=2$) of SAdS BHs are governed by a second-order differential
equation for a master variable $\Psi$ \cite{Horowitz:1999jd,Cardoso:2001bb}:
\beq
&&f^2\frac{d^2\Psi}{dr^2}+ff'\frac{d\Psi}{dr}+\left(\omega^2-V_{l,s}\right )\Psi=0\,,\label{waveeq}\\
&&V_{l,s}=f{\biggl[}\frac{l(l+1)}{r^2}+(1-s^2)\left (\frac{2M}{r^3}+\frac{4-s^2}{2L^2}\right ){\biggr]}\label{pot}\,,
\eeq
where $f=r^2/L^2+1-r_0/r$, $L$ is the AdS radius, $r_0$ is related to the
horizon radius $r_+$ through $r_0/L=(r_{+}/L)^{3}+r_{+}/L$, and we assume that
the perturbations depend on time as $e^{-i\omega t}$. As usual, we define a
``tortoise'' coordinate $r_*$ by the relation $dr/dr_*=f$ (so that $r_*
\to-\infty$ as $r\to r_+$) .

\begin{figure}[ht]
\begin{center}
\begin{tabular}{c}
\epsfig{file=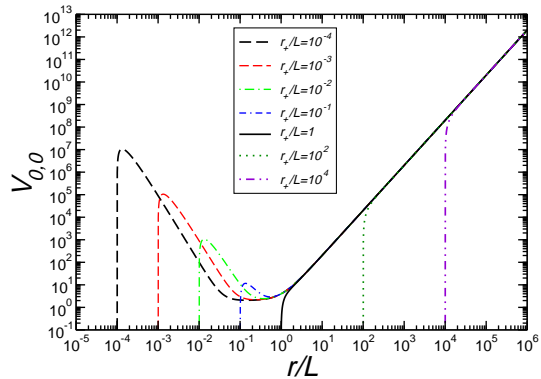,width=6cm,angle=270} 
\end{tabular}
\caption{Potential for scalar field ($s=0$) perturbations of a SAdS background
  with $l=0$. Different lines refer to different values of $r_+/L$. A
  potential well develops for small BHs ($r_+/L<1$).
  \label{fig:potsads}}
\end{center}
\end{figure}
The potential for scalar-field perturbations of SAdS BHs is shown in
Fig.~\ref{fig:potsads} for different values of $r_+/L$, ranging from ``large''
BHs with $r_+/L\sim 10^4$ to ``small'' BHs with $r_+/L\sim 10^{-4}$. Notice
how a potential well of increasing depth and width develops in the small BH
limit ($r_+/L\ll 1$).

Close to the horizon, where the potential $V_{l,s} \to 0$, we require
ingoing-wave boundary conditions:
\be \Psi \sim e^{-i\omega r_*}\,,\quad 
r_* \to-\infty\,~(r\to r_+)
\label{bchorizon} \,.\ee
Near spatial infinity ($r\to \infty$) the asymptotic behavior is
\be
\Psi_{s=0} \sim Ar^{-2}+Br\,,\quad \Psi_{s=1,~2} \sim A/r+B\,.\label{BCs}
\ee
Regular scalar-field perturbations should have $B=0$, corresponding to
Dirichlet boundary conditions at infinity. The case for electromagnetic and
gravitational perturbations is less clear, and there are indications that
Robin boundary conditions may be more appropriate in the context of the
AdS/CFT correspondence
\cite{Moss:2001ga,Michalogiorgakis:2006jc,Bakas:2008gz,Rocha:2008fe}. With
this caveat, most calculations in the literature assume Dirichlet boundary
conditions,
so we choose to work with those.

In general, a solution with the correct boundary conditions at infinity
behaves near the horizon ($r_*\to -\infty$) as
\be \Psi \sim A_{\rm in}e^{-i\omega r_*}+A_{\rm out} e^{i\omega r_*} \sim
\alpha \cos \omega r_*+\beta \sin \omega r_*\,,\label{bchorizon2}\ee
with $\alpha=A_{\rm out}+A_{\rm in}\,,\,\beta=i(A_{\rm out}-A_{\rm in})$.
For increased numerical accuracy, in our calculations we use a higher-order
expansion of the form
\be 
A_{\rm in}(1+a(r-r_+))e^{-i\omega r_*}+A_{\rm out} (1+a^{*}(r-r_+)) e^{i\omega
  r_*}\,,
\ee
with
\be 
a=\frac{2l(l+1)+(s^2-1)\left((s^2-6)r_+^2/L^2-2\right)}
{(2r_+/L)(1+3r_+^2/L^2-2i\omega r_+)}\,.
\ee

The problem is analogous to axial gravitational-wave scattering by compact
stars, as long as we replace the ``outgoing-wave boundary condition at
infinity'' in the stellar case by an ``ingoing-wave boundary condition at the
horizon'' in the SAdS case (compare our Fig.~\ref{fig:potsads} with Fig.~1 in
Ref.~\cite{ChandrasekharFerrari}). Quasi-bound states for the potential
(\ref{pot}) should show up as Breit-Wigner resonances of the form
(\ref{breitwigner}) for {\it real} $\omega$.

\begin{figure}[ht]
\begin{center}
\begin{tabular}{c}
\epsfig{file=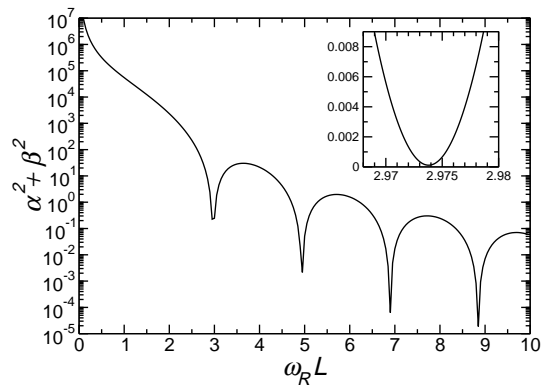,width=6cm,angle=270} 
\end{tabular}
\caption{A plot of $\alpha^2+\beta^2$ for scalar field SAdS perturbations with
  $l=0$, $r_+/L=10^{-2}$. Resonances are seen when $\omega_R\simeq 3+2n$,
  i.e. close to the resonant frequencies of the pure AdS spacetime. In the
  inset we show the behavior near the minimum, which allows us to extract the
  decay time by a parabolic fit. \label{fig:breit}}
\end{center}
\end{figure}
Fig. \ref{fig:breit} shows a typical plot of $\alpha^2+\beta^2$ as a function
of $\omega_R L$.  The pronounced dips correspond to the location of a
resonance, $\omega=\omega_R$, and the inset shows a zoomed-in view of one such
particular resonance. Once a minimum in $\alpha^2+\beta^2$ is located, the
imaginary part $\omega_I$ can be found by a parabolic fit of
$\alpha^2+\beta^2$ to the Breit-Wigner expression (\ref{breitwigner}).
Alternative (but equivalent) expressions are $\omega_I=-\beta
/\alpha'=\alpha/\beta'$, where a prime stands for the derivative with respect
to $\omega_R$, evaluated at the minimum
\cite{ChandrasekharFerrari,Chandrasekhar:1992ey}. We used these three
different expressions to estimate numerical errors in the computed quasinormal
frequencies. In the next subsections we briefly report on our results for
$s=0$ and $s=2$, respectively.

\subsection{Scalar field perturbations}

The series solution method presented by Horowitz and Hubeny
\cite{Horowitz:1999jd} was used by Konoplya in Ref.~\cite{Konoplya:2002zu} to
compute quasinormal frequencies of small SAdS BHs. The series has very poor
convergence properties for $r_+/L<1$, and QNM calculations in this regime take
considerable computational time. As seen in Fig.~\ref{fig:potsads}, the
potential for small SAdS BHs is able to sustain quasi-bound states, so we
expect the resonance method to be well adapted to the study of small BHs.

\vskip 1mm
\begin{table}[ht]
\caption{\label{tab:smallsads} The fundamental $l=0$ QNM frequencies for small
  SAdS BHs for selected values of $r_+/L$. The series solution data is taken
  from Ref.~\cite{Konoplya:2002zu}. In the table, $\delta \omega_R\,L\equiv
  3-\omega_R\,L$.}
\begin{tabular}{ccccc}  \hline
\multicolumn{1}{c}{} & \multicolumn{2}{c}{\rm series}& \multicolumn{2}{c}{ \rm resonance}\\
$L/r_+$ & $\delta \omega_R\,L^2/r_+$   & $\omega_I \,L^3/r_+^2$& $\delta \omega_R\,L^2/r_+ $   & $\omega_I\, L^3/r_+^2$  \\ 
\hline
\hline
12    & 3.064 &9.533         & 2.992 &9.662  \\ 
20    & 2.922 &7.720         & 2.912&7.840 \\ 
30    & 2.805 &6.660         & 2.802&6.714   \\ 
50    &       &                &2.700&5.952\\ 
100   &       &                & 2.610 &5.471\\
200  &        &                &2.580  &5.266 \\
%
%
%
500  &      &                &2.560 &5.158  \\
1000&      &                 &2.554        &5.125\\
2000&     &                  &2.550      &5.109\\
5000&     &                   &2.550    &5.100\\
\hline
\hline
\end{tabular}
\end{table}
\vskip 1mm
In Table \ref{tab:smallsads} we list QNMs for $l=0$ scalar field perturbations
and for different BH sizes, comparing (where possible) results from the
resonance method with Konoplya's series expansion calculation.
A cubic fit of our data for $L/r_+>30$ yields
$\omega_I\,L=5.00r_+^2/L^2+47.70r_+^3/L^3$, a quartic fit yields
$\omega_I=5.09(r_+/L)^2+33.59(r_+/L)^3+485.09(r_+/L)^4$, and fits with higher
order terms basically leave $a$ and $b$ unchanged with respect to the quartic
fit.
The numerical results are consistent with Horowitz and Hubeny's prediction
that $\omega_I\propto r_+^2$, and they are in very good agreement with
analytic predictions by Cardoso and Dias \cite{Cardoso:2004hs}, who derived a
general expression for the resonant frequencies of small BHs in AdS for
$M\omega_R \ll 1$ regime. Setting $a=0$ in Eq.~(33) of
Ref.~\cite{Cardoso:2004hs}, their result is
\beq
L\,\omega=l+3+2n-i\omega_I L\,,
 \label{carddias}
\eeq
where $n$ is a non-negative integer and
\beq
\label{deltaparameter}
& &\omega_I\,L\simeq -\gamma_0 \left[ \left ( l+3+2n\right )\,  
\left(r_+/L\right)^{\,2l+2}\right]/\pi\,,\\
\nn
& & \gamma_0\equiv \frac{2^{-1-6l}\,(l!)^2\,\Gamma\left[-l-1/2\right]^2\Gamma\left[5+2l+2n\right]}
{(3+2n)(3+2l+2n)\Gamma\left[l+1/2\right]^2\Gamma\left[2+2n\right]}\,.
\eeq
For $l=n=0$ one gets $\omega_I=16(r_+/L)^2/\pi\sim 5.09(r_+/L)^2$, in excellent
agreement with the fits. For general $l$ Eq.~(\ref{carddias}) predicts an
$r_+^{2+2l}$ dependence, in agreement with our results for $l=0,~1,~2$.
Moreover we find excellent agreement with the proportionality coefficient
predicted by Eq.~(\ref{deltaparameter}). Higher overtones are also well
described by Eqs.~(\ref{carddias}) and (\ref{deltaparameter}).

Our results show that the resonant frequency $\omega_R$ always approaches the
pure AdS value in the small BH limit, generally confirming the analysis by
Grain and Barrau \cite{Grain:2006dg}. However our numerics disagree with Grain
and Barrau's semi-classical calculation of the monopole mode ($l=0$). We find
that all modes including the monopole reduce to pure AdS in the small BH
limit. More precisely, as $r_+/L \to 0$ we find
\be
\omega_R\,L=l+3+2n-k_{ln}\,r_+/L\,,\quad n=0,\,1,\,2,\,\dots
\ee
with $k_{l0}\sim 2.6,~1.7,~1.3$ for $l=0,~1,~2\,$, respectively.

\begin{figure}[ht]
\begin{center}
\begin{tabular}{c}
\epsfig{file=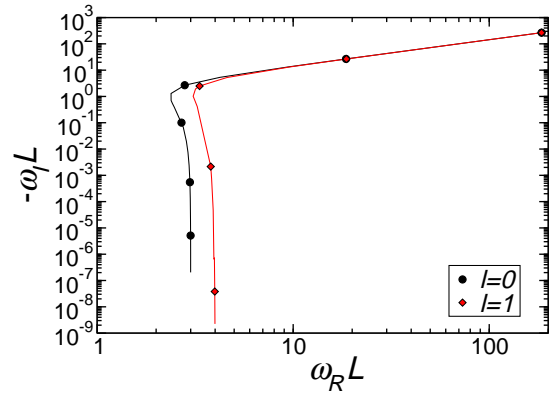,width=6cm,angle=270} 
\end{tabular}
\caption{Track traced in the complex plane ($\omega_R L,\omega_I L$) by the
  fundamental $l=0,1$ scalar field QNM frequencies as we vary the BH size
  $r_+/L$. Counterclockwise along these tracks we mark by circles and diamonds
  the frequencies corresponding to decreasing decades in $r_+/L$
  ($r_+/L=10^2,10^1,10^0,10^{-1},...$). \label{fig:SAdSsize}}
\end{center}
\end{figure}

Our results for scalar field perturbations are visually summarized in
Fig.~\ref{fig:SAdSsize}, where we combine results from the resonance method
and from the series solution (compare e.g. Fig.~1 of
Ref.~\cite{Berti:2003ud}). Modes with different $l$ coalesce in the large BH
regime (top-right in the plot), as long as $l\ll r_+/L$. As shown in
Fig.~\ref{fig:potsads}, the potential for small SAdS BHs develops a well
capable of sustaining quasi-stationary, long-lived modes. It should not be
surprising that small and large BH QNMs have such a qualitatively different
behavior.

\subsection{Gravitational perturbations}

We have also searched for the modes of Regge-Wheeler or vector-type
gravitational perturbations. The potential (\ref{pot}) for gravitational
perturbations of small BHs does not develop a local minimum. Nevertheless it
develops a local maximum which, when imposing Dirichlet boundary conditions
[see the discussion following Eq.~(\ref{BCs})] can sustain quasi-bound
states. For this reason we expect the resonance method to be useful also in
this case.

For small BHs our numerical results agree with the following analytic
estimate, derived under the assumption that $M\omega_R \ll 1$
\cite{Cardoso:2006wa}:
\beq
\omega_I\,L\simeq -\gamma_2 \left (
l+2+2n\right )\,\left(r_+/L\right)^{\,2l+2}
 \,,\label{deltas2}
\eeq
with
\begin{widetext}
\beq
& & \gamma_2\equiv \frac{(l+1)(2+l+2n)\,\Gamma\left[1/2-l\right]\Gamma\left[l\right]^3\Gamma\left[3+l\right]\Gamma\left[2(2+l+n)\right]}
{(l-1)^2(2n)!\Gamma\left[l+3/2\right]\Gamma\left[2l+1\right]\Gamma\left[2l+2\right]\Gamma\left[-1/2-l-2n\right]\Gamma\left[7/2+l+2n\right]}\,.
\label{deltaparameters2}
\eeq
\end{widetext}
For $n=0$ and $l=2$ this implies $\omega_I=1024(r_+/L)^6/45\pi \simeq
7.24(r_+/L)^6$, while a fit of the numerics yields $\omega_I\sim 7.44
(r_+/L)^6$.
Again, $\omega_R\,L$ approaches the pure AdS value
in the limit $r_+/L \to 0$:
\be
\omega_R\,L=l+2+2n-k_{ln}\,r_+/L\,,\quad n=0,\,1,\,2,\,\dots
\ee
For the fundamental $l=2$ mode we find $k_{20}\sim 1.4$.

\section{\label{eikonal}Long-lived modes in the eikonal limit}

A recent study of the eikonal limit ($l\gg1$) of SAdS BHs suggests that very
long-lived modes should exist in this regime \cite{Festuccia:2008zx}. Define
$r_b>r_c$ to be the two real zeros (turning points) of
$\omega_R^2-p^2f/r^2=0$. Then the real part of a class of long-lived modes in
four spacetime dimensions is given by the WKB condition
\be
2\int_{r_b}^{\infty}\frac{\sqrt{r^2\omega_R^2-p^2f}}{rf}\,dr=\pi\left (2n+5/2\right)\,,\label{bohrsommerfeld}
\ee
where $p=l+1/2$.  Their imaginary part is given by
\be
\omega_I=\frac{\gamma \Gamma}{8\omega_R}\,,\quad
\log \Gamma=2i\int_{r_b}^{r_c}\frac{\sqrt{r^2\omega_R^2-p^2f}}{rf}\,dr\,.
\ee
The prefactor $\gamma$, not shown in
Ref.~\cite{Festuccia:2008zx}, can be obtained by standard methods
\cite{Gamow:1928zz,Gurvitz:1988zz} with the result
\beq
\gamma&=&\left [\int_{r_b}^{\infty}\frac{\cos^2 \chi}{\sqrt{\omega_R^2-p^2f/r^2}}\frac{dr}{f}\right]^{-1}\,,\\
\nn
\chi&\equiv&\int_{r}^{\infty}\sqrt{\omega_R^2-p^2f/r^2}\,\frac{dr}{f}\,-\frac{\pi}{4}\,.
\eeq

\vskip 1mm
\begin{table}[ht]
\caption{\label{tab:longlived} The QNM frequencies for a $r_+/L=0.1$ BH for
  selected values of $l$, and comparison with the corresponding WKB
  prediction. 
}
\begin{tabular}{ccccc}  \hline
\multicolumn{1}{c}{} & \multicolumn{2}{c}{\rm WKB}& \multicolumn{2}{c}{ \rm resonance}\\
$(l,n)$ & $\omega_RL$   & $\log\omega_IL$& $\omega_RL$   & $\log\omega_IL$  \\ 
\hline 
\hline
3,0    & 5.8668& -16.40           & 5.8734 & -17.03  \\ 
3,1    & 7.6727& -12.61          & 7.6776 &-12.97 \\ 
3,2    & 9.4189& -9.40          & 9.4219 &-9.60   \\ 
%
%
4,0   &6.8830 & -22.13        & 6.8889    &-22.84\\ 
4,1   &8.7139& -18.30         & 8.7184&-18.41\\
4,2   &10.4960&-14.65         &10.4996& -14.72\\
5,0  &7.8945&-28.64        &7.8997&-28.76\\
5,1  &9.7426&-23.97        &9.7466&-24.02\\
5,2  &11.5482&-20.03        &11.5516&-20.06\\
\hline 
\hline
\end{tabular}
\end{table}
\vskip 1mm
The resonance method is well suited to analyze the eikonal limit, especially
for small BHs (for large BHs the existence of a dip in the potential well
requires very large values of $l$, which are numerically hard to deal
with). In Table \ref{tab:longlived} we compare the WKB results against
numerical results from the resonance method. Unfortunately machine precision
limitations do not allow us to extract extremely small imaginary parts. The
agreement with the WKB condition of Ref.~\cite{Festuccia:2008zx} is
remarkable, even for relatively small values of $l$. Our numerics conclusively
confirm the existence of very long-lived modes in the SAdS geometry, but the
numerical results for the damping timescales disagree by orders of magnitude
with the corresponding results by Grain and Barrau \cite{Grain:2006dg}. A
reanalysis of the assumptions implicit in their method would be useful to
understand the cause of this disagreement.


\section{Conclusions and outlook}
The method described here provides a reliable and accurate alternative to the
series solution method \cite{Horowitz:1999jd}, to be used in regimes where the
former has poor convergence properties.  Together, the two methods allow an
almost complete characterization of the spectrum of BHs in AdS backgrounds,
encompassing both small and large BHs.  As an application of the method, we
have explicitly confirmed the existence of the weakly damped modes predicted
by Refs.~\cite{Grain:2006dg,Festuccia:2008zx}.

Extensions of the resonance method to higher-dimensional
\cite{Horowitz:1999jd} and charged geometries \cite{Berti:2003ud,Wang:2004bv}
should be trivial. Our techniques may be useful to verify the existence of the
highly-real modes predicted by Ref.~\cite{Daghigh:2008jz}. Finally, it would
be interesting to investigate whether the resonance method described here is
of any use to investigate the eikonal limit of QNMs in asymptotically flat BH
spacetimes.

\section*{Acknowledgements}
We are indebted to Luis Lehner and Andrei Starinets for conversations that
started this project.  This work was partially funded by Funda\c c\~ao para a
Ci\^encia e Tecnologia (FCT) - Portugal through project
PTDC/FIS/64175/2006. P.P. thanks the CENTRA/IST for hospitality and financial
help and the Master and Back foundation programme of Regione Sardegna for a
grant. V.C. acknowledges financial support from the Fulbright Scholar
programme.


\end{document}